%% file: preprint.tex
\documentstyle[11pt,aaspp4,flushrt]{article}     

\lefthead{Salvador-Sol\'e, Solanes, \& Manrique}
\righthead{The Structure of Dark Matter Halos}

\newcommand{\der}{{\rm d}}

\begin{document}

\title{Merger vs. Accretion and the Structure of Dark Matter Halos} 
\author{Eduard Salvador-Sol\'e, Jos\'e Mar\'\i a Solanes, and Alberto Manrique}
\affil{Departament d'Astronomia i
Meteorologia, Universitat de Barcelona,\\ Avda. Diagonal 647, E-08028
Barcelona, Spain;\\ eduard@faess0.am.ub.es, solanes@pcess1.am.ub.es,
alberto@pcess2.am.ub.es}

\begin{abstract}

High-resolution N-body simulations of hierarchical clustering in a wide
variety of cosmogonies show that the density profiles of dark matter
halos are universal, with low mass halos being denser than their more
massive counterparts. This mass-density correlation is interpreted as
reflecting the earlier typical formation time of less massive
objects. We investigate this hypothesis in the light of formation times
defined as the epoch at which halos experience their last major
merger. Such halo formation times are calculated by means of a
modification of the extended Press \& Schechter formalism which
includes a phenomenological frontier, $\Delta_{\rm m}$, between tiny
and notable relative mass captures leading to the distinction between
merger and accretion. For $\Delta_{\rm m}\sim 0.6$, we confirm that the
characteristic density of halos is essentially proportional to the mean
density of the universe at their time of formation. Yet,
proportionality with respect to the critical density yields slightly
better results for open universes. In addition, we find that the scale
radius of halos is also essentially proportional to their virial radius
at the time of formation.

We show that these two relations are consistent with the following
simple scenario. Violent relaxation caused by mergers rearranges the
structure of halos leading to the same density profile with universal
values of the dimensionless characteristic density and scale
radius. Between mergers, halos grow gradually through the accretion of
surrounding layers by keeping their central parts steady and expanding
their virial radius as the critical density of the universe diminishes.

\end{abstract}


\section{Introduction}\label{intro}

High-resolution N-body simulations of hierarchical clustering in the
standard CDM cosmogony carried out by Navarro, Frenk, \& White
(1996\markcite{NFW96}) show that the spherically averaged
equilibrium density profiles of dark matter halos with masses ranging
from dwarf galaxy to rich cluster scales are well fitted (see however Moore
et al. 1997\markcite{Mo97}) by the formula
\begin{equation}
{\rho(x)\over{\rho_{\rm crit}}}=\delta_{\rm c} 
{x_{\rm s}^3\over{x(x+x_{\rm s})^2}}.\label{rho}
\end{equation}
In equation (\ref{rho}), $\rho_{\rm crit}$ is the critical density of
the universe, $x=r/R$ is the radius scaled to the so-called virial
radius $R$, and $\delta_{\rm c}=\rho_{\rm c}/\rho_{\rm crit}$ and
$x_{\rm s}=r_{\rm s}/R$ are two dimensionless parameters giving the
characteristic density and scale radius, respectively, of the density
profile. These latter two parameters are linked through the relation
\begin{equation}
\delta_{\rm c}={200\over 3}{x_{\rm s}^{-3}\over [\ln (1+x_{\rm
s}^{-1})-(1+x_{\rm s})^{-1}]},\label{relation}
\end{equation}
arising from the steadiness condition that the mean density within $R$
is equal to $200\times\rho_{\rm crit}$. Thus, the dimensional density
profile of a halo with mass $M$ at a given time $t$ (the latter two
quantities fixing the values of $R$ and $\rho_{\rm crit}$ in a given
cosmogony) is governed by one single free parameter. Note that the
inverse of $x_{\rm s}$ is a direct measure of the halo concentration.

More recently, Navarro, Frenk, \& White (1997, hereafter
NFW)\markcite{NFW}, and in independent work, Cole \& Lacey
(1997)\markcite{CL97} and Tormen, Bouchet, \& White
(1997),\markcite{TBW97} have shown that expression (\ref{rho}) provides
equally good fits to the density profile of dark halos in a number of
other cosmogonies, including flat and open models, with or without a 
cosmological constant, and with different initial power spectra of
Gaussian density fluctuations. In all the cosmogonies investigated the
parameter $\delta_{\rm c}$ has been found to correlate with mass in such a way that low
mass halos are denser than those of high mass. This mass-density
correlation is interpreted as reflecting the earlier typical formation
time of less massive objects. As shown by NFW, the correlation is well
described by a simple model in which the characteristic density
$\rho_{\rm c}$ of a halo of present mass $M_0$ is proportional to the
mean density of the universe at the corresponding formation redshift
$z_{\rm f}$, or equivalently,
\begin{equation}
\delta_{\rm c}=C\Omega_0[1+z_{\rm f}(M_0)]^3.\label{fit}
\end{equation}
To compute $z_{\rm f}(M_0)$, NFW used the expression derived by Lacey
\& Cole (1993,\markcite{LC93} LC) in the framework of the Press \&
Schechter (1974,\markcite{PS74} PS) prescription for the cumulative
probability that the mass of a halo following single $M(t)$ tracks
reduces to some fraction of its present mass, $f$, taken by NFW as a
free parameter. They find that for $f\le 0.01$ the predicted typical
mass-density relations fit all their simulations reasonably well for
essentially the same proportionality factor $C$. Although this result
strictly refers to present-day halos, it should also apply to halos at
any redshift for scale-free cosmogonies and those in which the
evolution of structure is close to being self-similar.

In spite of this remarkable result one cannot overlook the fact that
the distribution of formation times based on single $M(t)$ tracks is
not fully adequate for estimating the time at which a parent halo
reaches, for the first time, a fraction $f$ of its present mass
(cf. LC\markcite{LC93}). Moreover, the fact that $f$ must be less than
or equal to $0.01$ poses two problems. Firstly, it leads to an
ambiguous definition of the formation time, since a progenitor with
$M<0.5M_0$ is not necessarily along the main lineage. Second, it is
difficult to understand how the present structure of a halo can bear
any relationship to the epoch in which some progenitor reached such a
small fraction of the current halo mass. More importantly, the
definition of the formation time in the LC clustering model does not
distinguish between notable mass increases occurring more or less
abruptly in time. Major deviations from equilibrium and subsequent
violent relaxation take place only when halos of comparable masses
merge, while tiny mass captures have a negligible effect on the
structure of the capturing systems. Numerical simulations of
hierarchical clustering (Cole \& Lacey 1997\markcite{CL97}; Thomas et
al. 1997\markcite{Th97}) indeed show that halos with no evidence of a
recent major merger are in steady state within $R$, despite the fact
that they are continually capturing small halos.

Kitayama \& Suto (1996)\markcite{KS96} have attempted to describe the
formation and destruction of halos within the extended PS prescription
by differentiating between notable and tiny relative mass
captures. Their model lacks, however, a consistent definition for the
formation of halos because all halo captures involved in the same
common merger are counted separately as giving rise to different new
halos. A similar, but fully consistent approach, has been followed
independently by Manrique \& Salvador-Sol\'e (1995, 1996, hereafter
MS95 and MS96). These authors \markcite{MS95}\markcite{MS96} have
developed a semi-analytical clustering model within the framework of
the peak formalism, hereafter referred to as the CUSP (Confluent System
of Peak trajectories) model, which distinguishes naturally between
major and minor mergers, hereafter simply referred to as (true) mergers
and accretion. This allows one to define unambiguously the halo
formation and destruction times corresponding, respectively, to their
last and next merger. Unfortunately, to be fully satisfactory the CUSP
model requires a more accurate expression for the peak-peak correlation
at small separations than is presently available (Manrique et al. 1997,
M\&CO)\markcite{Ma97}.

In the present paper, we develop a self-consistent modification of the
LC model which, drawing inspiration from the CUSP model, differentiates
merger from accretion. This model, which retains the simplicity and
good predictive properties of the original model (Lacey \& Cole
1994\markcite{LC94}) while including better motivated formation and
destruction time estimates, is used to investigate the origin of the
empirical mass-density and related mass-radius correlations, as well as
their implications for the evolution of halo structure in hierarchical
cosmogonies. The paper is organized as follows. The modified LC model
is presented in \S\ \ref{exten}. It is applied in \S\ \ref{corr} to the
study of the empirical mass-density and mass-radius correlations. The
results of this analysis are summarized in \S\ \ref{diss}.

\section{Merger vs. Accretion and the PS Formalism}\label{exten}

A halo survives as long as it evolves by accretion, or equivalently, as
long as it captures only relatively tiny systems. Otherwise, it merges,
which automatically leads to its destruction. Note that when a halo is
captured by one that is more massive it merges and is destroyed in the
event, but the capturing halo may survive provided that the captured
mass is relatively small. Only those events in which {\it all\/} the
initial halos merge and are destroyed give rise to the formation of new
halos.

The preceding definitions do not affect the abundance of halos at a
given time, only the description of their growth. It is, therefore, not
surprising that the CUSP model also distinguishing between merger and
accretion predicts a halo mass function (MS95)\markcite{MS95} that is
highly similar to the PS one as in the LC model. Accordingly, in the
modified LC clustering model we propose, the mass function is 
equal to the PS mass function
\begin{equation}
N(M,t)=N_{\rm LC}(M,t)=\left ({2\over \pi}\right )^{1/2}{\rho_0\over
M^2}{\delta_{\rm coll}(t)\over \sigma(M)}\left |{\der\ln\sigma\over \der\ln
M}\right | \exp\left [-{\delta_{\rm coll}(t)^2\over 2\sigma^2(M)}\right
],
\label{mf} 
\end{equation}
where $\rho_0$ is the present mean mass density of the universe,
$\delta_{\rm coll}(t)$ is the critical overdensity for collapse at $t$
linearly extrapolated to the present time, and $\sigma(M)$ is the
current r.m.s. overdensity on spheres encompassing a mass $M$.

The instantaneous merger rate for halos of mass $M$ at $t$ per
infinitesimal range of final masses $M'>M$, or specific merger rate,
predicted by the CUSP model (MS96)\markcite{MS96} is also close to the
corresponding rate predicted by the original LC model, down to some
value $\Delta_{\rm m}$ of the relative captured mass $\Delta M/M\equiv
(M'-M)/M$, where it shows a sharp cutoff. This cutoff reflects the fact
that, in the CUSP model, captures of small mass halos relative to $M$
are not computed as mergers, but simply contribute to accretion. In
contrast, the specific merger rate predicted by the LC model
\begin{eqnarray}
r^{\rm m}_{\rm LC}(M\rightarrow M',t)=\left ({2\over \pi}\right
)^{1/2}\left |{\der\delta_{\rm coll}\over \der t}\right
|{1\over\sigma^2(M')}\left |{\der\sigma(M')\over\der M'}\right|
{1\over[1-\sigma^2(M')/\sigma^2(M)]^{3/2}}\nonumber\\ \label{mr}\\
\times\exp\left[-{\delta_{\rm coll}(t)^2\over 2}\left
({1\over\sigma^2(M')}-{1\over\sigma^2(M)}\right )\right ]\nonumber,
\end{eqnarray}
keeps on increasing monotonically for small $\Delta M/M$ because any
mass capture is regarded, in this model, as a merger, and the number
density of small mass halos diverges. Following this result we modify
the original LC model by including a threshold $\Delta_{\rm m}$ in the
relative mass captured by a halo for such an event to be considered a
merger, smaller mass captures only contributing to continuous
accretion. With this modification the new specific merger rate takes
the form
\begin{equation}
r^{\rm m}(M\rightarrow M',t)=\cases{0&if $M<M'\le M(\Delta_{\rm
m}+1)$;\cr r^{\rm m}_{\rm LC}(M\rightarrow M',t)&if $M(\Delta_{\rm
m}+1)<M'$,\cr}\label{mr2}
\end{equation}
while the total mass increase rate for halos of mass $M$ at $t$,
$r_{\rm mass}(M,t)\equiv \der M/\der t$, splits into two
contributions, one arising from mergers, or mass merger rate,
\begin{equation}
r^{\rm m}_{\rm mass}(M,t)=\int_{M(\Delta_{\rm m}+1)}^\infty\,\Delta
M\,r^{\rm m}(M\rightarrow M',t)\,\der M',\label{mir}
\end{equation}
and the other arising from accretion, or mass accretion rate,
\begin{equation}
r^{\rm a}_{\rm mass}(M,t)=\int_M^{M(\Delta_{\rm m}+1)}\,\Delta
M\,r^{\rm m}_{\rm LC}(M\rightarrow M',t)\,\der M'.
\end{equation}

As shown by M\&CO\markcite{Ma97}, the mass function, the specific
merger rate, and the mass accretion rate determine the behavior of the
entire CUSP model. This is also the case for the modified LC
model. The specific merger rate determines the mass merger rate
(eq. [\ref{mir}]), as well as the destruction rate of halos with mass
$M$ at $t$
\begin{equation}
r^{\rm d}(M,t)=\int_{M(\Delta_{\rm m}+1)}^\infty r^{\rm m}(M\rightarrow
M',t)\,\der M'.
\label{dr}
\end{equation}
Likewise, the formation rate can be written as
\begin{equation}
r^{\rm f}[M(t),t]={\der\ln N[M(t),t]\over \der t}+r^{\rm d}[M(t),t]+
\partial_M r^{\rm a}_{\rm mass}(M,t)\bigl|_{M=M(t)},\label{conserv}
\end{equation}
from the conservation equation for the number density of halos per unit
mass along mean mass accretion tracks, $M(t)$, solution of the
differential equation
\begin{equation}
{\der M\over\der t}=r^{\rm a}_{\rm mass}[M(t),t]. \label{massr}
\end{equation} 

Finally, the distributions of formation and destruction times in the
modified LC model are given by expressions identical to those in the
CUSP model (see M\&CO). In particular, the distribution of formation
times for halos at $t_0$ with masses between $M_0$ and $M_0+\delta
M_0$, with $\delta M_0$ arbitrarily small, takes the form
\begin{equation}
\Phi_{\rm f}(t)\equiv{1\over N_{\rm pre}(t_0)}\,{\der N_{\rm pre}\over \der t}
=r^{\rm f}[M(t),t]\,
\exp\biggl\{-\int_t^{t_0} r^{\rm f}[M(t'),t']\,\der t'\biggr\},\label{Phif}
\end{equation}
with $M(t)$ the mass of these halos at $t$ calculated along their mean
mass accretion tracks. The median of this distribution is adopted as
the typical halo formation time.

Before concluding this section, we should clarify the fact that the
distinction adopted between merger and accretion is not motivated by
the results of $N$-body simulations, but obeys the desire to
differentiate schematically the dynamic effects on halo structure of
tiny and notable relative mass captures. Note also that while the
merger cutoff in the CUSP model {\it arises naturally\/} from the peak
ansatz and the assumed distinction between merger and accretion (see
MS96), the threshold for merger, $\Delta_{\rm m}$, in the present
modified LC model is {\it a free phenomenological parameter\/} which,
for simplicity, will be considered independent of $M$ and $t$ (one
assumption implies the other in scale-free universes). Strictly
speaking, the rearrangement of a halo after the merger of two
progenitors depends on the relative gain of energy per unit mass rather
than simply on the relative mass increase. However, as the former
quantity is largely determined by the latter, this simplifying
assumption is justified.

\section{The $\delta_{\rm c}(M_0)$ and $x_{\rm s}(M_0)$ Correlations}
\label{corr}

Next we investigate the possible origin of the mass-density and
mass-radius correlations exhibited by halos in hierarchical universes
in the light of the modified LC model developed in the preceding
section. To do this we use the numerical data of NFW, which comprises
the eight different cosmogonies listed in Table 1. The first column of
this Table lists the power spectra, while columns (2) and (3) list the
values of $\Omega_0$ and $\lambda_0\equiv\Lambda/ (3H_0^2)$,
respectively. We list in column (4) the present r.m.s. density
fluctuation in $8\,h^{-1}\,$Mpc spheres, $\sigma_8$. To facilitate the
comparison among the cosmogonies, masses are scaled to the values of
the present characteristic mass $M_\ast$, defined through
$\sigma(M_\ast)=\delta_{\rm coll}(t_0)$, which are listed in column (5)
of Table 1. All the models have $h=0.5$, except the $\Lambda$CDM model
which has $h=0.75$, with the Hubble constant defined as
$H_0=100\,h\,{\rm km\,s^{-1}Mpc^{-1}}$.

In Figure 1 we show the best fits (by a standard least squares
minimization in logarithmic units) to the empirical $\delta_{\rm
c}(M_0)$ correlation obtained using the fitting formula (\ref{fit}) for
three different values of $\Delta_{\rm m}$. The value 0.6 corresponds
to the best overall fit when $\Delta_{\rm m}$ is varied from 0.1 to 0.9
in steps of one tenth. This value is also favored individually by the
three flat power-law spectrum models with $n=-1$, $-0.5$, and $0$,
which are those that best discriminate among the different
predictions. The two open scale-free models favor a value of
$\Delta_{\rm m}=0.5$, while the remaining power spectrum model and the
two CDM models favor $\Delta_{\rm m}=0.7$, although marginally. In
other words, as it is apparent from Figure 1, a value of $\Delta_{\rm
m}$ $\sim 0.6$ gives reasonably good fits in each of the cosmogonies
investigated. In contrast, the predictions corresponding to the
extreme values 0.1 and 0.9 do not describe the numerical data well in
practically any case.

Given the formation time distribution function (eq. [\ref{Phif}]) and
relation (\ref{fit}) we can readily derive the distribution functions
of $\log (\delta_{\rm c})$ for any value of $M_0$. Figure 2 shows the
distributions obtained in the SCDM model for five different values of
$M_0$, other cosmogonies giving qualitatively similar results. They are
in good overall agreement with the empirical distributions of points:
the maxima are near to the values of $\log (\delta_{\rm c})$
corresponding to the median formation redshifts, and the spreads have
the right magnitude and show a trend to diminish with increasing
$M_0$. This indicates that relation (\ref{fit}) also applies to
individual halos and that their characteristic density, $\rho_{\rm c}$,
remains essentially equal to $C$ times the mean cosmic density at their
time of formation. Note that, according to the PS mass function, low
mass halos are severely underrepresented in the empirical samples with
respect to more massive ones, indicating that the selection of the
former has been much stricter. In this manner, earlier formation times
may have been artificially favored, since the older the halos, the
better they satisfy the requirement of having a relaxed
appearance. This might explain the slightly smaller scatter shown by
the empirical distributions for small mass halos. This effect and the
slight bias also introduced by our simple fitting procedure (we have
assumed constant, symmetrical errors) might affect to some extent the
quantitative results of the fits, but the previous conclusions should
prevail.

The values of $C$ listed in column (6) of Table 1 show a much wider
variation with the cosmogony (an overall factor 100) than in NFW (only
a factor two; see their Table 1). Although the possible biases
mentioned above might in part be responsible for this variation, the
marked departure from a hypothetical common value shown by the values
of $C$ in open cosmogonies seems real. We have investigated the
possibility of reducing the scatter in the $\Omega_0<1$ cases by
devising a slightly different model which has the added value of
providing a straightforward physical interpretation of the empirical
mass-density correlation. In the new model, the characteristic density,
$\rho_{\rm c}$, of halos with current mass $M_0$ is assumed to be
proportional to the {\it critical} density of the universe at their
time of formation, instead of to the mean cosmic density at that
time. With such a proportionality, not only does $\rho_{\rm c}$ remain
fixed from the time of halo formation, but also the initial value of
$\delta_{\rm c}$ is {\it universal} (i.e., independent of mass and time
in self-similar universes). From the form this fitting model adopts in
dimensionless units
\begin{equation}
\delta_{\rm c}=\delta_{\rm cf}{\Omega_0\over
\Omega[z_{\rm f}(M_0)]}\,[1+z_{\rm f} (M_0)]^3,\label{fit2}
\end{equation}
it is apparent that the value of $\delta_{\rm c}$ when halos form is
equal to the proportionality factor $\delta_{\rm cf}$.

The best overall fit of the empirical data with the model (\ref{fit2})
is obtained again for $\Delta_{\rm m}=0.6$. As can be seen from Figure
1, the fits in the open cases are slightly better than in the original
model (\ref{fit}), while the two models give, of course, identical
results in the $\Omega_0=1$ cases. The overall variation shown by the
proportionality factor $\delta_{\rm cf}$ in different cosmogonies has
diminished, although a trend of $\delta_{\rm cf}$ with cosmogony is
still present. We note that some theoretical studies predict a
dependence on the cosmogony of the typical halo density profiles
resulting from violent relaxation (e.g., Syer \& White 1997).

Relation (\ref{relation}) between the dimensionless parameters
$\delta_{\rm c}$ and $x_{\rm s}$ tells us that the value of $x_{\rm s}$
shown by halos at their time of formation, hereafter denoted by $x_{\rm
sf}$, is also universal. The values of $x_{\rm sf}$ inferred from those
of $\delta_{\rm cf}$ drawn from the previous fits are listed in column
(8) of Table 1. The universality of $x_{\rm sf}$ is equivalent to
stating that the dimensional scale radius $r_{\rm s}$ of halos at their
time of formation is proportional, with universal proportionality
factor equal to $x_{\rm sf}$, to their virial radius $R$ at that
epoch. This raises the question: is the scale radius $r_{\rm s}$ of
current halos also proportional, with {\it identical} proportionality
factor, to their virial radius $R$ at their time of formation? Or
equivalently, does the value of $r_{\rm s}$ for current halos of mass
$M_0$ coincide with the value this parameter had when they formed, 
as for $\rho_{\rm c}$? The answer to these questions is not trivial
since it depends on the relation between the initial and current halo
masses, that is, on the typical mass accreted since their formation.

The modified LC model allows us to estimate the mass accreted by
halos. Hence, we can readily check the previous proportionality, which
in dimensionless units takes the form
\begin{equation}
x_{\rm s}=x_{\rm sf}{R\{M[z_{\rm f}(M_0),M_0]\}\over
R(M_0)}.\label{fit3}
\end{equation}
In Figure 3 we show the results of directly fitting this model to the
$x_{\rm s}(M_0)$ empirical correlation. Once again we find the best
overall fit for $\Delta_{\rm m}=0.6$. More importantly, the best values
of $x_{\rm sf}$, listed in column (9) of Table 1, are in fairly good
agreement (just slightly larger on average) with those listed in column
(8). Note that, despite the convoluted calculations involved (given a
halo of current mass $M_0$ one must first calculate its formation
redshift, then, using the modified LC model, its mass at $z_{\rm f}$,
and finally, through the steadiness condition, the corresponding value
of $R$), the fits are as good as those obtained for the mass-density
correlation.

The agreement, case by case, between the two {\it independent\/} values
of $x_{\rm sf}$ given by the correlations $\delta_{\rm c}(M_0)$
(through the relation [\ref{relation}]) and $x_{\rm s}(M_0)$ (through our
clustering model) supports the overall validity of the modified
LC model for $\Delta_{\rm m}\sim 0.6$, and of the theoretical
relations (\ref{fit2}) and (\ref{fit3}). The physical interpretation of
the latter can be summarized as follows: {\it i\/}) the values of the
dimensionless parameters, $\delta_{\rm cf}$ and $x_{\rm sf}$,
characterizing the radially averaged density profiles of halos at
formation are universal, and {\it ii\/}) the values of the
corresponding dimensional parameters, $\rho_{\rm c}$ and $r_{\rm s}$,
remain fixed as long as halos evolve by accretion.

The fact that for a given halo the values of parameters $\rho_{\rm c}$
and $r_{\rm s}$ are set when it forms tells us that the only effect of
accretion is the gradual expansion of the halo virial radius $R$ in
order to permanently satisfy the steadiness condition. We have directly
tested this corollary by comparing the mass increase experienced by
halos since their formation predicted by the modified LC model, with
the mass increase that results from taking halos with a density profile
of the form (\ref{rho}), with fixed values of $\rho_{\rm c}$ and
$r_{\rm s}$, and progressively increasing the virial radius $R$ as
$\rho_{\rm crit}$ diminishes. As expected, we have found a fair degree
of agreement between the two mass evolutions, the maximum departure
being less than 35\% in any one cosmogony.

\section{Conclusions}\label{diss}

The $\delta_{\rm c}(M_0)$ correlations predicted by the modified LC
model for $\Delta_{\rm m}= 0.6$, assuming the relation (\ref{fit}),
match the empirical data as well as the NFW predictions for $f=
0.01$. We therefore confirm, with a more compelling formation time
estimate, the claim by these authors that the characteristic density
shown by dark halos in equilibrium is proportional to the mean density
of the universe at the time they form. We want to stress that while the
two different formation time estimates give similarly good fits, this
does not imply that the difference in their definitions is merely
formal: for any given halo mass, the typical formation redshifts used
by NFW are appreciably larger, typically by a factor of two, than those
obtained in the model presented here. We have also shown that the fits
for the open models can be improved further if one assumes instead a
proportionality with respect to the critical density of the universe at
halo formation.

The modified LC model presented in \S\ \ref{exten}, together with the
definitions of halo formation and destruction times with which it
deals, relies on the schematic differentiation of the dynamic effects
of tiny and notable mass captures. According to this scenario, the
structure of halos would be fixed through violent relaxation in the
last major merger which they had experienced, while between two
consecutive major mergers halos would remain essentially unaltered,
mass accretion only producing a progressive expansion of their envelope
as new surrounding layers fall in and relax through gentle phase
mixing. The results of our analysis in \S\ \ref{corr} agree with this
simplified description. To be more specific, we have found that the
empirical $\delta_{\rm c}(M_0)$ and $x_{\rm s}(M_0)$ correlations are
consistent with the fact that halos show, at formation, the same
density profile with universal values of $\delta_{\rm c}$ and $x_{\rm
s}$. Until a new major merger takes place, the density profile of halos
keeps essentially the same form, though the values of $\delta_{\rm c}$
and $x_{\rm s}$ shift as the dimensional characteristic density and
scale radius, $\rho_{\rm c}$ and $r_{\rm s}$, remain fixed while the
virial radius $R$ expands accordingly to the decrease of the cosmic
critical density. As shown for the SCDM case by Avila-Reese, Firmani,
\& Hern\'andez (1997)\markcite{Av97}, the latter evolution seems to be
a natural consequence of adiabatic-invariant secondary infall. On the
other hand, some effect along the lines proposed by Syer \& White
(1997) might explain the universal halo density profiles resulting from
major mergers.

\acknowledgments
\begin{sloppypar}
We thank J.F. Navarro for kindly providing the data for the empirical
correlations appearing in Figures 1 and 3. The present work has been
supported by the Direcci\'on General de Investigaci\'on Cient\'\i
fi\-ca y T\'ecnica under contract PB96-0173.
\end{sloppypar}

%
%
\newpage 

\begin{figure}
\plotone{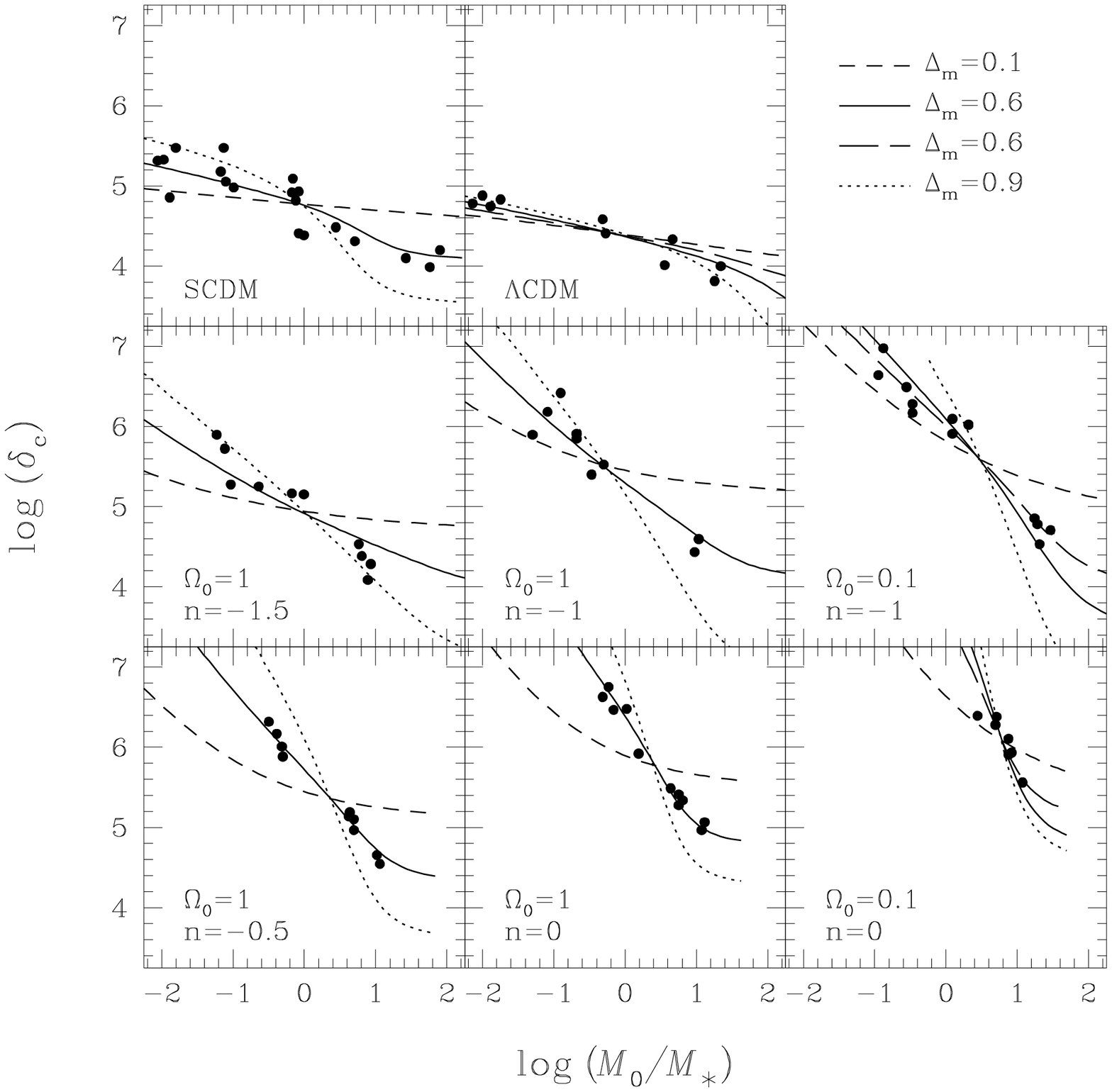}
\epsscale{1.0}
\caption[fig1.eps]
{Predicted $\delta_{\rm c}(M_0)$ correlations compared with the
empirical data from NFW's N-body simulations (filled circles). Dotted
and short-dashed curves show the predictions for two extreme values of
$\Delta_{\rm m}$, while the solid curves correspond to the value of
this parameter that gives the best overall fit. Cosmogonies with
$\Omega_0<1$ contain a fourth long-dashed curve which shows, for
$\Delta_{\rm m}=0.6$, the predictions arising from the assumption that
$\rho_{\rm c}$ is proportional to $\rho_{\rm crit}[z_{\rm f}(M_0)]$,
instead of to the mean density of the universe at halo formation.}
\end{figure}

\begin{figure}
\plotone{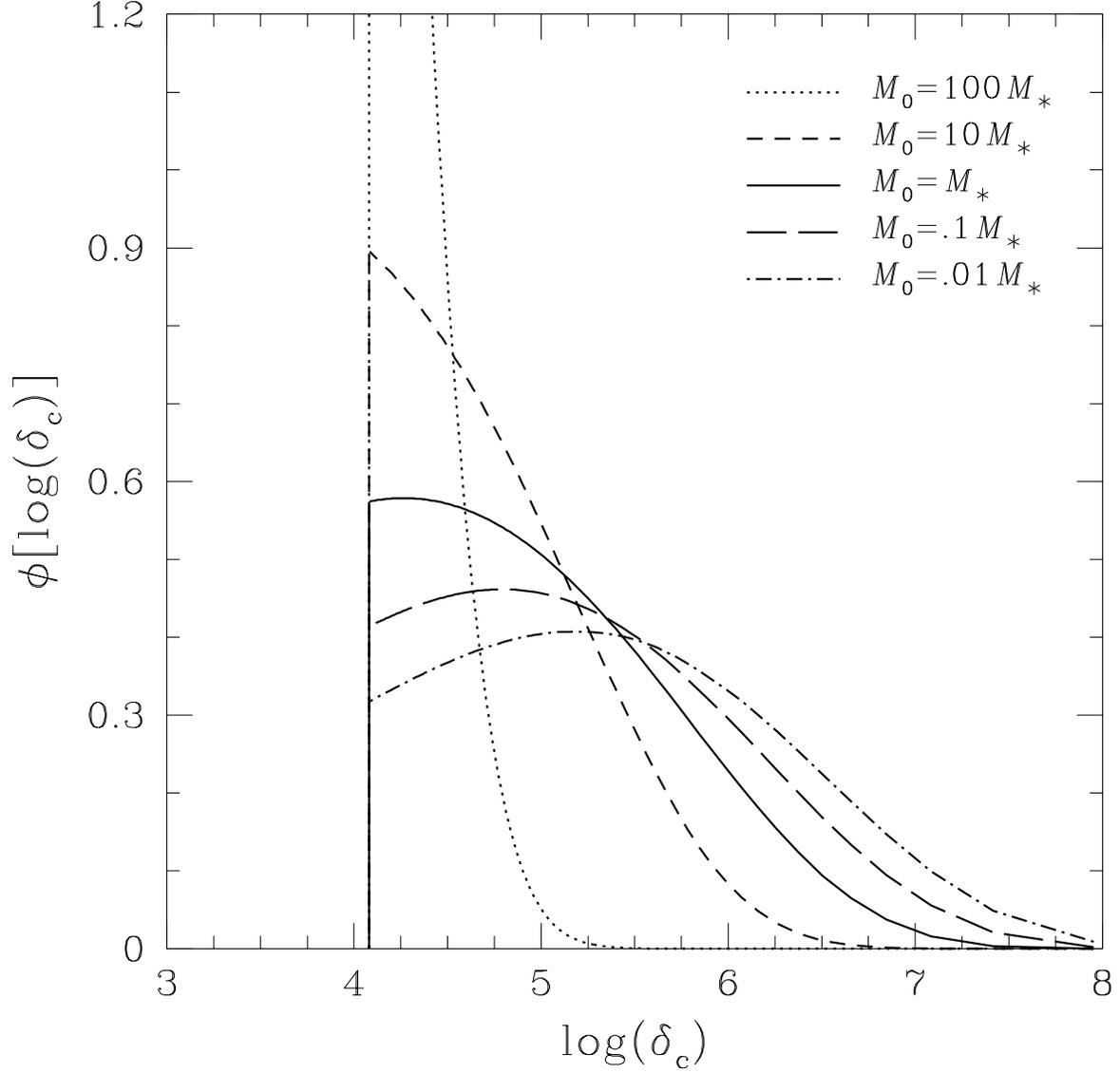}
\caption[fig2.eps]{Distributions of ${\rm log}
[\delta_{\rm c}(M_0)]$ in the SCDM cosmogony for halos of different
masses predicted by the modified LC model with $\Delta_{\rm m}=0.6$.}
\end{figure}

\begin{figure}
\plotone{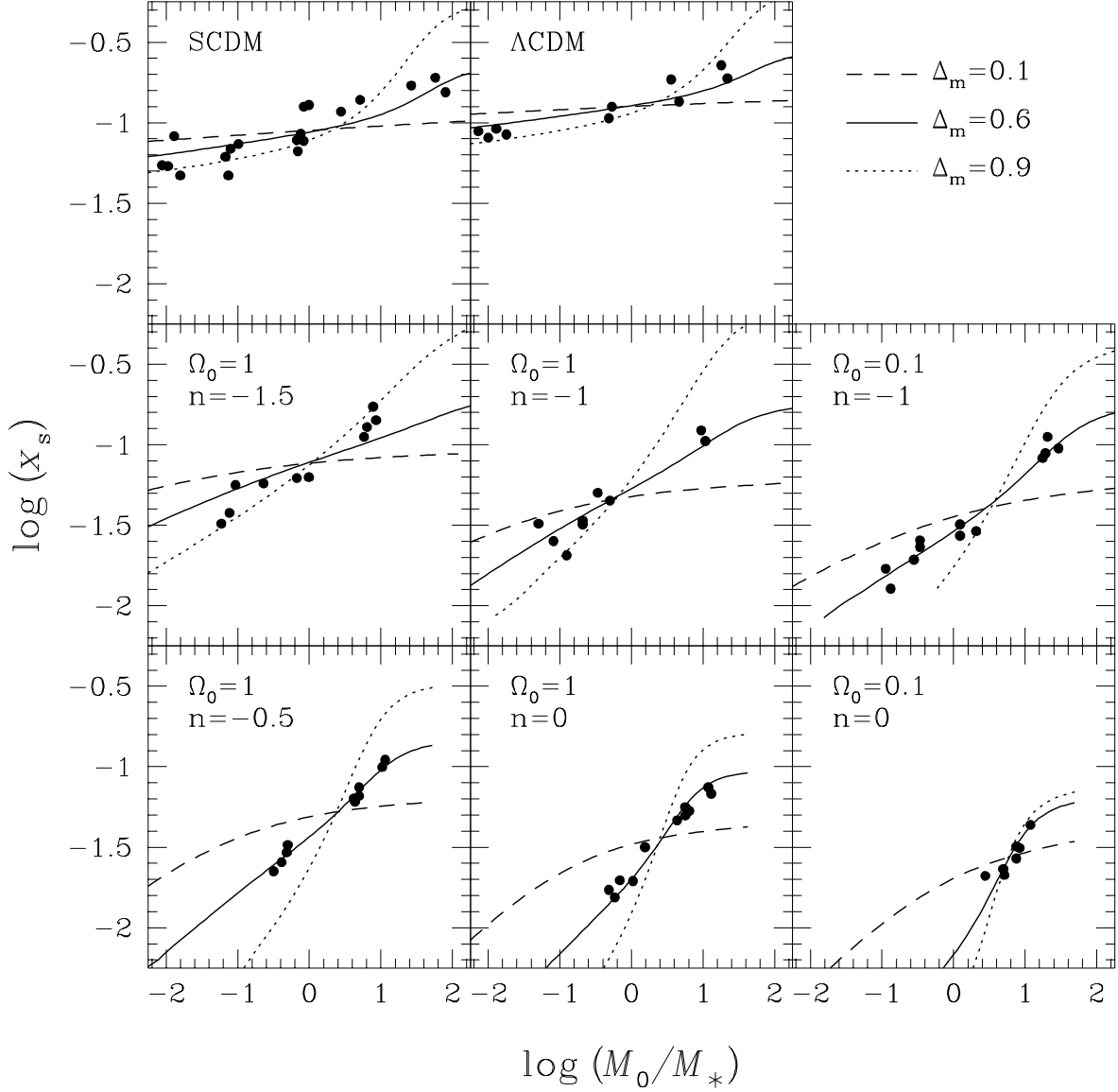}
\caption[fig3.eps]
{Predicted $x_{\rm s}(M_0)$ correlations compared with the empirical
data from NFW's N-body simulations (solid circles).}
\end{figure}

%
%
\newpage 
\include{table}

\end{document}

%% file: table.tex


\def\tablevspace#1{\noalign{\vskip#1}}

\begin{deluxetable}{lrrrccccc}
\tablewidth{0pt}
\tablenum{1}
\tablecaption{Parameters of the models}
\tablehead{
\colhead{$P(k)$}                  & \colhead{$\Omega_0$}       &   
\colhead{$\lambda_0$}             & \colhead{$\sigma_8$}       &
\colhead{$M_*/M_\odot$}           &\colhead{$C$}               &
\colhead{$\delta_{\rm cf}$}       &\colhead{$x_{\rm sf}$\tablenotemark{a}}  & 
\colhead{$x_{\rm sf}$}   
\nl \tablevspace{5pt}
\colhead{(1)}                  & \colhead{(2)}            &
\colhead{(3)}                  & \colhead{(4)}            &
\colhead{(5)}		       & \colhead{(6)}            &
\colhead{(7)}                  & \colhead{(8)}		  &
\colhead{(9)}}
\tablenotetext{a}{implied by $\delta_{\rm cf}$}
\startdata
SCDM & 1.0 & 0.0 & 0.63 & 3.08$\times 10^{13}$ & 1.21$\times 10^4$ & 1.21$\times 10^4$ & 0.173 & 0.229 \nl
$\Lambda$CDM & 0.25 & 0.75 & 1.3 & 6.31$\times 10^{13}$ & 4.21$\times 10^3$ & 3.77$\times 10^3$ & 0.291 & 0.285 \nl
$n=-1.5$ & 1.0 & 0.0 & 1.0 & 1.47$\times 10^{14}$ & 8.30$\times 10^3$ & 8.30$\times 10^3$ & 0.204 & 0.223 \nl
$n=-1.0$ & 1.0 & 0.0 & 1.0 & 2.48$\times 10^{14}$ & 1.28$\times 10^4$ & 1.28$\times 10^4$ & 0.169 & 0.181 \nl
         & 0.1 & 0.0 & 1.0 & 2.82$\times 10^{13}$ & 2.65$\times 10^4$ & 1.00$\times 10^4$ & 0.188 & 0.184 \nl
$n=-0.5$ & 1.0 & 0.0 & 1.0 & 3.40$\times 10^{14}$ & 2.16$\times 10^4$ & 2.16$\times 10^4$ & 0.135 & 0.148 \nl
$n= 0.0$ & 1.0 & 0.0 & 1.0 & 4.19$\times 10^{14}$ & 6.19$\times 10^4$ & 6.19$\times 10^4$ & 0.088 & 0.096 \nl
         & 0.1 & 0.0 & 1.0 & 4.56$\times 10^{13}$ & 5.77$\times 10^5$ & 1.33$\times 10^5$ & 0.064 & 0.065 \nl
\enddata
\end{deluxetable}